\documentclass[english]{revtex4-2}
\usepackage[T1]{fontenc}
\usepackage[latin9]{inputenc}
\usepackage{textcomp}
\usepackage{slashed}
\usepackage{amsmath}
\usepackage{amssymb}
\usepackage{graphicx}

\makeatletter

\newcommand{\lyxmathsym}[1]{\ifmmode\begingroup\def\b@ld{bold}
  \text{\ifx\math@version\b@ld\bfseries\fi#1}\endgroup\else#1\fi}

\usepackage{babel}
\def\CM{\mathrm{\scriptscriptstyle CM}}

\makeatother

\usepackage{babel}
\begin{document}
\title{Finite perturbation theory for the relativistic Coulomb problem}
\author{Scott E. Hoffmann}
\address{School of Mathematics and Physics,~~\\
 The University of Queensland,~~\\
 Brisbane, QLD 4072~~\\
 Australia}
\email{scott.hoffmann@uqconnect.edu.au}

\begin{abstract}
We present a novel form of relativistic quantum mechanics and demonstrate
how to solve it using a recently derived unitary perturbation theory,
within partial wave analysis. The theory is tested on a relativistic
problem, with two spinless, equal mass particles, in which the interaction
is entirely given by a Coulomb potential. As such, it is not meant
to reproduce experimental results for the scattering of two electrons,
but is intended as a test of our calculation methods. We find that
this perturbation theory gives finite results at second order. This
is unlike other versions of perturbation theory, which find divergent
results at second and all higher orders. We calculate differential
cross sections in the nonrelativistic regime, where we find excellent
agreement with the Rutherford formula. Then, well into the relativistic
regime, we find differential cross sections with similar shapes to
the Møller formula and differing from that formula by less than an
order of magnitude.
\end{abstract}
\maketitle

\section{Introduction}

There are two aims of this paper. The first is to apply a novel perturbation
theory to the problem of the Coulomb scattering of two (spinless)
charged particles. This problem has been treated before with other
forms of perturbation theory, with the well-known result that all
corrections at second and higher order diverge (\citet{Dalitz1951}).
The reason, in this context, is that certain integrals over particle
separation do not converge because the Coulomb potential falls off
insufficiently fast at large particle separation. We note that even
the first-order result is problematic (\citet{Collas2021}).

The new method used here (\citet{Hoffmann2021a}) involves working in
momentum space and a prescription to use principal part integration
on the momentum integrals. We will find that this procedure produces
finite results in agreement with expectations when applied to the
Coulomb potential.

The other aim of this paper is to show that some features of relativity
can be easily incorporated into our model. For our purposes, relativity
gives a different dependence of single-particle energy on momentum
from the nonrelativistic form, and can add new terms to the interaction.
For the latter, an example is the emergence of the spin-orbit coupling
term when the Foldy-Wouthuysen transformation is applied to the Dirac
equation (\citet{Foldy1950}). In our model, we will introduce the relativistic
form of the single-particle energy. We will consider an entirely Coulomb
potential in this spinless case.

We note that relativity constrains measurements made in different
frames, but does not tell us about measurements in one frame. The
position profile of a porcupine depends on how that animal has grown,
and varies from animal to animal. Relativity tells us only how a moving
frame would measure a changed porcupine profile. We will propose that
there must be a four-component potential operator that transforms
as a four-vector to have a covariant relativistic quantum mechanics.
We will find that in a centre of mass (CM) frame, the spatial components
must vanish. The form of the potential in a CM frame is arbitrary
for our considerations, but is determined by the model of microscopic
interactions in quantum electrodynamics (\citet{Weinberg2012}). Future
work will define these potentials and explore the consequences.

Throughout this paper, we use Heaviside-Lorentz units, in which $\hbar=c=\epsilon_{0}=\mu_{0}=1.$

\section{Finite, relativistic, perturbation theory}

\subsection{Relativistic Schrödinger equations}

We are motivated by the success of nonrelativistic quantum mechanics,
in which the total Hamiltonian is written as the sum of a free part
and a part responsible for interaction (a potential $\hat{V}$):
\begin{equation}
\hat{H}_{i}=\hat{H}_{f}+\hat{V}.\label{eq:2.1}
\end{equation}
Here $f$ indicates free, $i$ indicates interacting, hats indicate
operators.

Thus we propose for relativistic quantum mechanics the four equations
($\mu=0,1,2,3$)
\begin{equation}
\hat{P}_{i}^{\mu}=\hat{P}_{f}^{\mu}+\hat{V}^{\mu}.\label{eq:2.2}
\end{equation}
Here the potential of nonrelativistic quantum mechanics becomes the
zero component of what must transform as a four-vector, as do the
components of $P_{f}^{\mu}$ and $P_{i}^{\mu},$ to guarantee relativistic
covariance. This is
\begin{equation}
U^{\dagger}(L)\,V^{\mu}\,U(L)=L_{\phantom{\mu}\nu}^{\mu}\,V^{\nu},\label{eq:2.3}
\end{equation}
where $U(L)$ is the unitary representation of a general Lorentz transformation,
$L,$ and $L_{\phantom{\mu}\nu}^{\mu}$ is the corresponding coordinate
transformation matrix. For processes involving photons, we assert
that $V^{\mu}$ must be gauge invariant to arrive at physical results
independent of the gauge.

For the scattering of two spinless ``electrons'', we choose as free
basis vectors $|\,P,l,m;f\,\rangle.$ In the CM frame, these are defined
in terms of the individual momentum eigenvectors as
\begin{equation}
|\,(2\omega^{\CM},\boldsymbol{0}),l,m;f\,\rangle=\int d^{2}\hat{p}^{\CM}\,|\,+\boldsymbol{p}^{\CM},-\boldsymbol{p}^{\CM};f\,\rangle\,p^{\CM}\,Y_{lm}(\hat{\boldsymbol{p}}^{\CM}).\label{eq:2.4}
\end{equation}
Then a boost takes them to general four-momentum
\begin{equation}
|\,P,l,m;f\,\rangle=U(\Lambda(\frac{\boldsymbol{P}}{E}))\,|\,(2\omega^{\CM},0),l,m;f\,\rangle.\label{eq:2.5}
\end{equation}
We can ignore the total linear momentum, $\boldsymbol{P},$ by using
the normalization
\begin{equation}
\langle\,+\boldsymbol{p}_{1}^{\CM},-\boldsymbol{p}_{1}^{\CM};f\,|\,+\boldsymbol{p}_{2}^{\CM},-\boldsymbol{p}_{2}^{\CM};f\,\rangle=\delta^{3}(\boldsymbol{p}_{1}^{\CM}-\boldsymbol{p}_{2}^{\CM}),\label{eq:2.6}
\end{equation}
which has no consequences.

These are eigenvectors of total four-momentum $P^{\mu}=(E,\boldsymbol{P})^{\mu}.$
They are also eigenvectors of $\boldsymbol{L}^{2}$ and $L_{z},$
where $\boldsymbol{L}$ is the total angular momentum (which in this
case is entirely orbital angular momentum) in the CM frame (where
$\boldsymbol{P}=0$). In the more general case, where the two particles
can be any combination of electrons, positrons and photons, the basis
vectors take the form $|P,J,M,\lambda_{1},\lambda_{2};f\,\rangle,$
where $J$ and $M$ are the quantum numbers associated with the total
CM-frame angular momentum, $\boldsymbol{J},$ and $\lambda_{1}$ and
$\lambda_{2}$ are the CM-frame helicities of the two particles.

It is advantageous to work in momentum space, where the square roots
in the single-particle energies, $\omega(\boldsymbol{p})=\sqrt{\boldsymbol{p}^{2}+m_{e}^{2}},$
pose no problem. At this point there is no need to introduce spacetime
coordinates for the two particles. Note that to see a potential with
spacetime dependence requires taking matrix elements between two state
vectors, at least one of which has spacetime dependence.

Corresponding to each free basis vector is an interacting basis vector
that solves the Schrödinger equations (eq. (\ref{eq:2.2})). We will
explore this correspondence shortly.

We take matrix elements
\begin{equation}
\langle\,P_{f},l_{f},m_{f};f\,|\,\hat{P}_{i}^{\mu}\,|\,P_{i},l_{i},m_{i};i\,\rangle=\langle\,P_{f},l_{f},m_{f};f\,|\,\hat{P}_{f}^{\mu}+\hat{V}^{\mu}\,|\,P_{i},l_{i},m_{i};i\,\rangle,\label{eq:2.7}
\end{equation}
which reduce to
\begin{align}
(P_{f}^{\mu}-P_{i}^{\mu})\,\langle\,P_{f},l_{f},m_{f};f\,|\,P_{i},l_{i},m_{i};i\,\rangle & =-\langle\,P_{f},l_{f},m_{f};f\,|\,\hat{V}^{\mu}\,|\,P_{i},l_{i},m_{i};i\,\rangle\nonumber \\
 & =-\int d^{4}P_{f}^{\prime}\sum_{l_{f}^{\prime}=0}^{\infty}\sum_{m_{f}^{\prime}=-l_{f}^{\prime}}^{l_{f}^{\prime}}\langle\,P_{f},l_{f},m_{f};f\,|\,\hat{V}^{\mu}\,|\,P_{f}^{\prime},l_{f}^{\prime},m_{f}^{\prime};f\,\rangle\langle\,P_{f}^{\prime},l_{f}^{\prime},m_{f}^{\prime};f\,|\,P_{i},l_{i},m_{i};i\,\rangle,\label{eq:2.8}
\end{align}
a set of integral equations. We ensure that the free basis vectors
are orthonormalized and complete so that we can insert the completeness
relation as we have just done.

\subsection{Simplification in the centre of mass frame}

A simplification is possible. We will find below that it is possible
to define a unitary transformation that relates free and interacting
basis vectors with the same quantum numbers: 
\begin{equation}
|\,P,l,m;i\,\rangle=U_{if}\,|\,P,l,m;f\,\rangle.\label{eq:2.9}
\end{equation}
Note that in the quantum electrodynamics formalism (\citet{Weinberg2012}),
the electron is given a free mass different from its mass after including
electromagnetic interactions. But the difference of these masses is
calculated to diverge and only the interacting mass is observable.
As we will see, the correspondence proposed here produces no divergences.

The transformation of operators that gives the same expectation values
is
\begin{equation}
\langle\,P,l,m;i\,|\,P_{i}^{\mu}\,|\,P,l,m;i\,\rangle=\langle\,P,l,m;f\,|\,P_{f}^{\mu}\,|\,P,l,m;f\,\rangle\Rightarrow P_{i}^{\mu}=U_{if}\,P_{f}^{\mu}\,U_{if}^{\dagger}.\label{eq:2.10}
\end{equation}

Note that we will have similar relations for the transformation of
rotation generators ($\boldsymbol{J}$) and boost generators ($\boldsymbol{K}$):
\begin{equation}
J_{i}^{k}=U_{if}\,J_{f}^{k}\,U_{if}^{\dagger}\quad\mathrm{and}\quad K_{i}^{k}=U_{if}\,K_{f}^{k}\,U_{if}^{\dagger}\quad\mathrm{for}\ k=1,2,3.\label{eq:2.11}
\end{equation}

Consequently, if
\begin{equation}
\hat{\boldsymbol{P}}_{f}\,|\,P,l,m;f\,\rangle=\boldsymbol{0},\label{eq:2.12}
\end{equation}
then
\begin{equation}
\hat{\boldsymbol{P}}_{i}\,|\,P,l,m;i\,\rangle=U_{if}\,\hat{\boldsymbol{P}}_{f}\,U_{if}^{\dagger}\,U_{if}\,|\,P,l,m;f\,\rangle=U_{if}\,\hat{\boldsymbol{P}}_{f}\,|\,P,l,m;f\,\rangle=\boldsymbol{0}.\label{eq:2.13}
\end{equation}
So, from eq. (), we see the vanishing of the matrix elements of the
three-vector part of the potential, $\hat{\boldsymbol{V}},$ in the
one CM frame that is the same in the free and interacting theories:
\begin{equation}
\langle\,(E_{f},\boldsymbol{0}),l_{f},m_{f};f\,|\,\hat{\boldsymbol{V}}\,|\,(E_{i},\boldsymbol{0}),l_{i},m_{i};i\,\rangle=\boldsymbol{0}.\label{eq:2.14}
\end{equation}

In what follows, we will perform calculations only in the CM frame,
knowing that we could transform differential cross sections to other
frames, if necessary.

\subsection{Translational and rotational invariance}

We define $\hat{V}=\hat{V}^{0}.$ We know
\begin{equation}
\langle\,P_{f},l_{f},m_{f};f\,|\,P_{i},l_{i},m_{i};i\,\rangle\propto\delta^{3}(\boldsymbol{P}_{f}-\boldsymbol{P}_{i})\label{eq:2.15}
\end{equation}
by translational invariance, so we can remove all reference to the
three-momenta and arrive at the simplified equation in the CM frame
(understanding $\boldsymbol{P}_{f}=\boldsymbol{P}_{i}=0$ throughout)
\begin{equation}
(E_{f}-E_{i})\,\langle\,E_{f},l_{f},m_{f};f\,|\,E_{i},l_{i},m_{i};i\,\rangle=-\int_{2m_{e}}^{\infty}dE_{f}^{\prime}\sum_{l_{f}^{\prime}=0}^{\infty}\sum_{m_{f}^{\prime}=-l_{f}^{\prime}}^{l_{f}^{\prime}}\langle\,E_{f},l_{f},m_{f};f\,|\,\hat{V}\,|\,E_{f}^{\prime},l_{f}^{\prime},m_{f}^{\prime};f\,\rangle\langle\,E_{f}^{\prime},l_{f}^{\prime},m_{f}^{\prime};f\,|\,E_{i},l_{i},m_{i};i\,\rangle.\label{eq:2.16}
\end{equation}

We are considering the case of a potential that is rotationally invariant
($[\boldsymbol{J}_{i},\hat{V}]=0$). Thus, by the Wigner-Eckhart theorem
(\citet{Messiah1961}), the matrix elements take the form
\begin{equation}
\langle\,E_{f},l_{f},m_{f};f\,|\,\hat{V}\,|\,E_{f}^{\prime},l_{f}^{\prime},m_{f}^{\prime};f\,\rangle\propto\delta_{l_{f}l_{f}^{\prime}}\delta_{m_{f}m_{f}^{\prime}}.\label{eq:2.17}
\end{equation}
Furthermore, if the potential is rotationally invariant, we expect
that the free to interacting transformation will also be rotationally
invariant, preserving the quantum numbers $l$ and $m$. A further
reduction, ignoring $m,$ gives
\begin{equation}
(E_{f}-E_{i})\,\langle\,E_{f},l;f\,|\,E_{i},l;i\,\rangle=-\int_{2m_{e}}^{\infty}dE_{f}^{\prime}\langle\,E_{f},l;f\,|\,\hat{V}\,|\,E_{f}^{\prime},l;f\,\rangle\langle\,E_{f}^{\prime},l;f\,|\,E_{i},l;i\,\rangle.\label{eq:2.18}
\end{equation}

\subsection{Eigenvectors of particle separation}

The phase shifts that we want to calculate come from the asymptotic
behaviour of interacting position wavefunctions as the separation,
$r,$ of the two particles increases without bound. Thus we need to
construct eigenvectors of the operator associated with this separation.
Note that we cannot here use the techniques of nonrelativistic quantum
mechanics, where the kinetic energy depends quadratically on the momentum,
to form the total linear momentum and the relative momentum.

We consider two different individual particle translations applied
to a free individual momentum eigenvector:
\begin{equation}
U(T_{1}(+\boldsymbol{r}/2))\,U(T_{2}(-\boldsymbol{r}/2))\,|\,\frac{1}{2}\boldsymbol{P}+\boldsymbol{p},\frac{1}{2}\boldsymbol{P}-\boldsymbol{p};f\,\rangle=|\,\frac{1}{2}\boldsymbol{P}+\boldsymbol{p},\frac{1}{2}\boldsymbol{P}-\boldsymbol{p}\,\rangle\,e^{-i\boldsymbol{p}\cdot\boldsymbol{r}}.\label{eq:2.19}
\end{equation}
This transformation changes the separation of the two particles by
$\boldsymbol{r}$ but leaves the average position unchanged. So we
define simultaneous eigenvectors of separation and total three-momentum
as
\begin{equation}
|\,\boldsymbol{r},\boldsymbol{P};f\,\rangle=\int\frac{d^{3}p}{(2\pi)^{\frac{3}{2}}}\,|\,\frac{1}{2}\boldsymbol{P}+\boldsymbol{p},\frac{1}{2}\boldsymbol{P}-\boldsymbol{p};f\,\rangle\,e^{-i\boldsymbol{p}\cdot\boldsymbol{r}},\label{eq:2.20}
\end{equation}
noting that
\begin{equation}
\hat{\boldsymbol{r}}=i\frac{\partial}{\partial\boldsymbol{p}}=i\frac{\partial}{\partial\boldsymbol{p}_{1}}-i\frac{\partial}{\partial\boldsymbol{p}_{2}}\label{eq:2.21}
\end{equation}
and $\hat{\boldsymbol{P}}$ commute.

Then we form eigenvectors of the magnitude of the separation and of
orbital angular momentum,
\begin{equation}
|\,r,l,m,\boldsymbol{P};f\,\rangle=\int d^{2}\hat{r}\,|\,\boldsymbol{r},\boldsymbol{P};f\,\rangle\,r\,Y_{lm}(\hat{\boldsymbol{r}}),\label{eq:2.22}
\end{equation}
where the argument of the spherical harmonic is here a unit vector,
not an operator. We note that $\hat{\boldsymbol{L}}=\hat{\boldsymbol{r}}\times\hat{\boldsymbol{p}}$
defined in the CM frame also commutes with $\boldsymbol{P}.$

Note that we will only be requiring these basis vectors for the special
case $\boldsymbol{P}=\boldsymbol{0},$ where $r$ is the separation
and $l,m$ label the orbital angular momentum, both in the CM frame.
Note also that these basis vectors will have calculable but very complicated
Lorentz transformation properties, as with any position eigenvectors.

If we choose the orthonormalization
\begin{equation}
\langle\,\frac{1}{2}\boldsymbol{P}_{a}+\boldsymbol{p}_{a},\frac{1}{2}\boldsymbol{P}_{a}-\boldsymbol{p}_{a};f\,|\,\frac{1}{2}\boldsymbol{P}_{b}+\boldsymbol{p}_{b},\frac{1}{2}\boldsymbol{P}_{b}-\boldsymbol{p}_{b};f\,\rangle=\delta^{3}(\boldsymbol{P}_{a}-\boldsymbol{P}_{b})\,\delta^{3}(\boldsymbol{p}_{a}-\boldsymbol{p}_{b}),\label{eq:2.23}
\end{equation}
then we find
\begin{equation}
\langle\,r_{a},l_{a},m_{a},\boldsymbol{P}_{a};f\,|\,r_{b},l_{b},m_{b},\boldsymbol{P}_{b};f\,\rangle=\delta^{3}(\boldsymbol{P}_{a}-\boldsymbol{P}_{b})\,\delta(r_{a}-r_{b})\,\delta_{l_{a}l_{b}}\delta_{m_{a}m_{b}}.\label{eq:2.24}
\end{equation}

We find the relation between eigenvectors of momentum magnitude and
eigenvectors of separation magnitude:
\begin{equation}
\langle\,r,l_{a},m_{a},\boldsymbol{P}_{a};f\,|\,p,l_{b},m_{b},\boldsymbol{P}_{b};f\,\rangle=\delta^{3}(\boldsymbol{P}_{a}-\boldsymbol{P}_{b})\,\delta_{l_{a}l_{b}}\delta_{m_{a}m_{b}}\,\sqrt{\frac{2}{\pi}}\,pr\,j_{l_{a}}(pr)=\delta^{3}(\boldsymbol{P}_{a}-\boldsymbol{P}_{b})\,\delta_{l_{a}l_{b}}\delta_{m_{a}m_{b}}\,y_{l_{a}}^{(f)}(r,p),\label{eq:2.25}
\end{equation}
in terms of the free spherical waves, $y_{l_{a}}^{(f)}(r,p),$ familiar
from partial wave analysis (\citet{Messiah1961}). This motivates our
definition for the interacting case,
\begin{equation}
\langle\,r,l_{a},m_{a},P_{a};f\,|\,p,l_{b},m_{b},P_{b};i\,\rangle\equiv\delta^{3}(\boldsymbol{P}_{a}-\boldsymbol{P}_{b})\,\delta_{l_{a}l_{b}}\delta_{m_{a}m_{b}}\,y_{l_{a}}^{(i)}(r,p),\label{eq:2.26}
\end{equation}
in terms of the interacting wavefunctions, $y_{l_{a}}^{(i)}(r,p).$

Note that the position eigenvectors are common to the free and interacting
theories. We do not define interacting position eigenvectors.

\subsection{Unitary perturbation theory}

A unitary transformation connecting the free and interacting theories,
in the nonrelativistic case, was introduced in \citet{Hoffmann2021a},
and a perturbation series was developed to second order in the coupling
strength. Only minor changes are needed to treat the relativistic
case.

In the CM frame, the total energies are
\begin{equation}
E(p)=2\sqrt{p^{2}+m_{e}^{2}},\label{eq:2.27}
\end{equation}
where $p$ is the magnitude of both single-particle momenta. Then
a change of quantum numbers to eq. (\ref{eq:2.18}) gives
\begin{equation}
(E(k)-E(p))\,\langle\,k,l;f\,|\,p,l;i\,\rangle=-\int_{0}^{\infty}dk^{\prime}\,\langle\,k,l;f\,|\,\hat{V}\,|\,k^{\prime},l;f\,\rangle\langle\,k^{\prime},l;f\,|\,p,l;i\,\rangle.\label{eq:2.28}
\end{equation}
The normalizations are
\begin{equation}
\langle\,E_{1},l;f/i\,|\,E_{2},l;f/i\,\rangle=\delta(E_{1}-E_{2})\Leftrightarrow\langle\,p_{1},l;f/i\,|\,p_{2},l;f/i\,\rangle=\delta(p_{1}-p_{2}).\label{eq:2.29}
\end{equation}

We expand the unitary transformation to second order in $\alpha$:
\begin{equation}
U_{if}=e^{-i\Theta}\cong1-i\alpha\,\Theta^{(1)}-\frac{i}{2}\alpha^{2}\Theta^{(2)}-\frac{1}{2}\alpha^{2}\Theta^{(1)2},\label{eq:2.30}
\end{equation}
where
\begin{equation}
\Theta\cong\alpha\,\Theta^{(1)}+\frac{1}{2}\alpha^{2}\,\Theta^{(2)}.\label{eq:2.31}
\end{equation}

Then we solve the equation
\begin{equation}
H_{i}+V=U_{if}\,H_{f}\,U_{if}^{\dagger}\label{eq:2.32}
\end{equation}
order by order, with $V=\mathcal{O}(\alpha)$. The results are
\begin{equation}
i[H_{f},\alpha\,\Theta^{(1)}]=V\quad\mathrm{and}\quad[H_{f},\alpha^{2}\,\Theta^{(2)}]=[\alpha\,\Theta^{(1)},V].\label{eq:2.33}
\end{equation}

The solutions for the matrix elements of $\Theta^{(1)}$ and $\Theta^{(2)}$
are
\begin{equation}
\langle\,k_{1},l;f\,|\,\alpha\,\Theta^{(1)}\,|\,k_{2},l;f\,\rangle=-iP\frac{\mathcal{V}_{l}(k_{1},k_{2})}{E(k_{1})-E(k_{2})}\label{eq:2.34}
\end{equation}
and
\begin{equation}
\langle\,k_{1},l;f\,|\,\alpha^{2}\,\Theta^{(2)}\,|\,k_{2},l;f\,\rangle=-iP\,\frac{1}{E(k_{1})-E(k_{2})}\int_{0}^{\infty}dk^{\prime}\,\left\{ \frac{\mathcal{V}_{l}(k_{1},k^{\prime})\mathcal{V}_{l}(k^{\prime},k_{2})}{E(k_{1})-E(k^{\prime})}-\frac{\mathcal{V}_{l}(k_{1},k^{\prime})\mathcal{V}_{l}(k^{\prime},k_{2})}{E(k^{\prime})-E(k_{2})}\right\} ,\label{eq:2.35}
\end{equation}
where
\begin{equation}
\mathcal{V}_{l}(k_{1},k_{2})=\langle\,k_{1},l;f\,|\,V\,|\,k_{2},l;f\,\rangle.\label{eq:2.36}
\end{equation}
Also
\begin{equation}
\langle\,k_{1},l;f\,|\,\alpha^{2}\,\Theta^{(1)2}\,|\,k_{2},l;f\,\rangle=-P\int_{0}^{\infty}dk^{\prime}\,\frac{\mathcal{V}_{l}(k_{1},k^{\prime})}{E(k_{1})-E(k^{\prime})}\frac{\mathcal{V}_{l}(k^{\prime},k_{2})}{E(k^{\prime})-E(k_{2})}.\label{eq:2.37}
\end{equation}
The symbol $P$ indicates that when these expressions are integrated
over momentum, the principal part prescription (also known as finding
the Cauchy principal value) must be used. We have seen in (\citet{Hoffmann2021a})
that use of this prescription eliminates divergences and give results
in agreement with an exactly solvable model.

Then the interacting position separation wavefunction to second order
is
\begin{align}
y_{l}^{(2)}(r,p) & =\langle\,r,l;f\,|\,U_{if}^{(2)}\,|\,p,l;f\,\rangle\nonumber \\
 & =y_{l}^{(f)}(r,p)-P\int_{0}^{\infty}dk\,y_{l}^{(f)}(r,k)\,\frac{\mathcal{V}_{l}(k,p)}{E(k)-E(p)}+\frac{1}{2}P\int_{0}^{\infty}dk\,y_{l}^{(f)}(r,k)\int_{0}^{\infty}dk^{\prime}\,\frac{\mathcal{V}_{l}(k,k^{\prime})\mathcal{V}_{l}(k^{\prime},p)}{(E(k)-E(k^{\prime}))(E(k^{\prime})-E(p))}\nonumber \\
 & \quad-\frac{1}{2}P\int_{0}^{\infty}dk\,y_{l}^{(f)}(r,k)\int_{0}^{\infty}dk^{\prime}\,\left\{ \frac{\mathcal{V}_{l}(k,k^{\prime})\mathcal{V}_{l}(k^{\prime},p)}{(E(k)-E(p))(E(k)-E(k^{\prime}))}-\frac{\mathcal{V}_{l}(k,k^{\prime})\mathcal{V}_{l}(k^{\prime},p)}{(E(k)-E(p))(E(k^{\prime})-E(p))}\right\} .\label{eq:2.38}
\end{align}

\subsection{Calculation of phase shifts}

For the Coulomb potential in the CM frame, the matrix elements are
\begin{align}
\mathcal{V}_{l}(k_{1},k_{2}) & =\int_{0}^{\infty}dr\,y_{l}^{(f)}(r,k_{1})\,\frac{\alpha}{r}\,y_{l}^{(f)}(r,k_{2})\nonumber \\
 & =\frac{\alpha}{\sqrt{\pi}}\,\frac{l!}{\Gamma(l+\frac{3}{2})}\,\begin{cases}
\left(\rho\right)^{l+1}\,_{2}F_{1}(\frac{1}{2},l+1;l+\frac{3}{2};\rho^{2}) & \rho<1,\\
\left(\frac{1}{\rho}\right)^{l+1}\,_{2}F_{1}(\frac{1}{2},l+1;l+\frac{3}{2};\frac{1}{\rho^{2}}) & \rho>1,
\end{cases}\label{eq:2.39}
\end{align}
with $\rho=k_{1}/k_{2}.$ Both sections of this function are singular
at $k_{1}=k_{2}.$ Close to the singularity, with $k_{1}=k_{2}(1+x)$
for small $x,$ we find the approximation (\citet{DLMF2020}, their
eq. (15.8.10))
\begin{equation}
\mathcal{V}_{l}(k_{2}(1+x),k_{2})\rightarrow\frac{\alpha}{\pi}(C_{l}-\ln|x|)+\mathcal{O}(x),\label{eq:2.40}
\end{equation}
with
\begin{equation}
C_{l}=\ln2-\boldsymbol{C}-\psi(l+1)\label{eq:2.41}
\end{equation}
and
\begin{equation}
\boldsymbol{C}=0.577215\dots\label{eq:2.42}
\end{equation}
known as Euler's constant. Here $\psi(z)$ is the logarithmic derivative
of the gamma function (\citet{Gradsteyn1980}, their section 8.36).

These functions $\mathcal{V}_{l}(k_{1},k_{2})$ have the property
that they have a singular point but are integrable across the singularity.
This will be sufficient for the principal part integrals to converge
to finite values.

To find the asymptotic form of $y_{l}^{(2)}(r,p)$ in eq. (\ref{eq:2.38}),
we first use the asymptotic forms of the free spherical waves (\citet{Messiah1961})
\begin{equation}
y_{l}^{(f)}(r,k)\rightarrow\sqrt{\frac{2}{\pi}}\,\sin(kr-l\frac{\pi}{2}).\label{eq:2.43}
\end{equation}

For the first-order term (with $k=p(1+x)$)
\begin{equation}
I^{(1)}\rightarrow-P\int_{-1}^{\infty}dx\,\sqrt{\frac{2}{\pi}}\,\{\sin(pr-l\frac{\pi}{2})\cos(prx)+\cos(pr-l\frac{\pi}{2})\sin(prx)\}\,\frac{\mathcal{V}_{l}(p(1+x),p)}{x}\left\{ \frac{E(p(1+x))+E(p)}{4p(2+x)}\right\} .\label{eq:2.44}
\end{equation}
We review the rules we learned from the similar case in \citet{Hoffmann2021a}:
\begin{itemize}
\item For the $\cos(prx)$ and $\sin(prx)$ integrals, the contributions
from $x>1$ will vanish asymptotically like $1/pr,$ since the integrand
has no singularities on that region and the sinusoids oscillate rapidly.
\item The factor $\sin(prx)/x\rightarrow\pi\,\delta(x)$ acts as a delta
function. When combined with the singularity of $\mathcal{V}_{l}(p(1+x),p),$
we have (with $z=prx$)
\begin{equation}
-\frac{\alpha}{\pi}\int_{-\infty}^{\infty}dz\,\mathrm{sinc}(z)(C_{l}-\ln|z/pr|)=\alpha(-C_{l}-\ln(pr)-\boldsymbol{C})=\alpha(-\ln(2pr)+\psi(l+1)).\label{eq:2.45}
\end{equation}
\item There is no delta function in the integral containing $\cos(prx).$
On $x\in[-1,1],$ only integrands even in $x$ will survive the principal
part integration. In this integrand, there will be factors of the
form $[A]_{-}[B]_{+}/x+[A]_{+}[B]_{-}/x,$ with $A=\mathcal{V}_{l}(p(1+x),p)$
and $B=\{E(p(1+x))+E(p)\}/p(2+x)$ and $[f(x)]_{\pm}=\frac{1}{2}\{f(x)\pm f(-x)\}.$
We note that the first of these terms and $[B]_{-}/x$ are finite
at the origin. So we consider integrals of the form
\begin{equation}
\int dx\,\ln x\,\cos(prx)=x\,\ln x\,\mathrm{sinc}(prx)-\frac{\mathrm{Si}(prx)}{pr},\label{eq:2.46}
\end{equation}
where $\mathrm{Si}(x)=\int_{0}^{x}dt\,\mathrm{sinc}\,t.$ This integral
vanishes at $x=0$ and decreases like $1/pr$ for $x\neq0.$
\end{itemize}
Thus we find (with $B(0)=E(p)/2p=1/2\beta,$ where $\beta=p/\sqrt{p^{2}+m_{e}^{2}}$
is the relativistic velocity magnitude for either of the particles)
\begin{equation}
I^{(1)}=\sqrt{\frac{2}{\pi}}\,\cos(pr-l\frac{\pi}{2})\,\eta\,\{-\ln(2pr)+\psi(l+1)\}\equiv\sqrt{\frac{2}{\pi}}\,\cos(pr-l\frac{\pi}{2})\,\{\delta_{l}^{(1)}(p)\},\label{eq:2.47}
\end{equation}
with
\begin{equation}
\delta_{l}^{(1)}(p)=-\eta\ln(2pr)+\eta\,\psi(l+1).\label{eq:2.48}
\end{equation}

Here $\eta=\alpha/2\beta$ corresponds to the parameter defined in
the nonrelativistic case, $\eta_{\mathrm{NR}}=\alpha/(p/\mu),$ with
$\mu=m_{e}/2$ the reduced mass in this case. We will find that our
perturbation series is in powers of $\eta$ rather than of $\alpha.$

We use similar methods for the second-order terms, employing the four
rules
\begin{align}
\int_{0}^{\infty}dk\,\frac{\sin(kr-l\frac{\pi}{2})}{E(k)-E(p)}\,f(k) & \rightarrow\cos(pr-l\frac{\pi}{2})\,\frac{\pi}{2\beta}\,f(p),\nonumber \\
\int_{0}^{\infty}dx\,\frac{\cos(kr-l\frac{\pi}{2})}{E(k)-E(p)}\,f(k) & \rightarrow-\sin(pr-l\frac{\pi}{2})\,\frac{\pi}{2\beta}\,f(p),\nonumber \\
\int_{-1}^{\infty}dx\,\frac{\sin(kr-l\frac{\pi}{2})}{E(k)-E(p)}\,\mathcal{V}_{l}(p(1+x),p)\,f(x) & \rightarrow\cos(pr-l\frac{\pi}{2})\,\delta_{l}^{(1)}(p)\,f(0)\nonumber \\
\int_{-1}^{\infty}dx\,\frac{\cos(kr-l\frac{\pi}{2})}{E(k)-E(p)}\,\mathcal{V}_{l}(p(1+x),p)\,f(x) & \rightarrow-\sin(pr-l\frac{\pi}{2})\,\delta_{l}^{(1)}(p)\,f(0),\label{eq:2.49}
\end{align}
all for $f(k)$ having no singularities. We evaluate the terms in
eq. (\ref{eq:2.38}) as written, performing each $k^{\prime}$ integral
before the $k$ integral. We find that changing the order of the double
momentum integrals changes some of the terms, but the total is invariant.
An ambiguity remains. Using partial fractions to cancel some terms
gives a different final result. We suspect this is because doing this
changes the set of poles in the integrands. This matter is under further
investigation.

We find the result
\begin{equation}
y_{l}^{(2)}(r,p)\rightarrow\sqrt{\frac{2}{\pi}}\,\sin(pr-l\frac{\pi}{2})\,\left\{ -\frac{1}{2}\delta_{l}^{(1)}(p)^{2}\right\} +\sqrt{\frac{2}{\pi}}\,\cos(pr-l\frac{\pi}{2})\,\left\{ \delta_{l}^{(1)}(p)+\frac{1}{2}\,\frac{\pi}{2\beta}\,P\int_{0}^{\infty}dk^{\prime}\,\frac{\mathcal{V}_{l}(k^{\prime},p)^{2}}{E(k^{\prime})-E(p)}\right\} .\label{eq:2.50}
\end{equation}
We note that the term proportional to $\sin(pr-l\frac{\pi}{2})$ has
the correct form required by unitarity at this order. The last term
is
\begin{equation}
\delta_{l}^{(2)}(p)=\frac{1}{2}\,\frac{\pi}{2\beta}\,P\int_{0}^{\infty}dk^{\prime}\,\frac{\mathcal{V}_{l}(k^{\prime},p)^{2}}{E(k^{\prime})-E(p)}=\frac{1}{2}\,\frac{\pi}{(2\beta)^{2}}\,P\int_{-1}^{\infty}dx\,\frac{\mathcal{V}_{l}(p(1+x),p)^{2}}{x}\,g(p,x),\label{eq:2.51}
\end{equation}
where
\begin{equation}
g(p,x)=2\beta\,\frac{E(p(1+x))+E(p)}{4p(2+x)}\rightarrow1\label{eq:2.52}
\end{equation}
at $x=0.$

\section{Calculation of differential cross sections}

We use the results of \citet{Hoffmann2017a}, where we considered
the scattering of a single-particle wavepacket from a Coulomb potential.
A slight modification of the differential cross section formula, here
in the CM frame, with the resulting form shown here, was necessary
to treat the scattering of two particles,
\begin{equation}
\frac{d\sigma}{d\Omega}=\frac{1}{4p^{2}}\left|\sum_{l=0}^{\infty}(2l+1)\,e^{-\epsilon^{2}(l+\frac{1}{2})^{2}}\,e^{i2(\bar{\delta}_{l}^{(1)}(p)+\delta_{l}^{(2)}(p))}\,e^{-(\delta-\Delta_{l}(p))^{2}/4}\,P_{l}(\cos\theta)\right|^{2}.\label{eq:3.1}
\end{equation}

The wavepacket treatment makes the sum in eq. (\ref{eq:3.1}) converge,
even for $\theta=0.$ It was previously thought that partial wave
analysis was not applicable to the Coulomb potential, as the sum over
$l$ in eq. (\ref{eq:3.1}) diverges for all scattering angles if
the convergence factor $\exp(-\epsilon^{2}(l+\frac{1}{2})^{2})$ is
not present. This treatment introduces another physical parameter,
\begin{equation}
\epsilon=\frac{\sigma_{p}}{p},\label{eq:3.2}
\end{equation}
where $\sigma_{p}$ is the standard deviation of the Gaussian probability
density in momentum space of all the initial and final wavepackets.
The standard deviation in position space at $t=0$ is $\sigma_{x},$
with $\sigma_{x}\sigma_{p}=\frac{1}{2}.$ In practice, we choose $\epsilon=0.001.$
In the nonrelativistic case, we found generally excellent agreement
with the Rutherford formula (\citet{Rutherford1911}),
\[
\frac{d\sigma}{d\Omega}_{\mathrm{Rutherford}}=\frac{m_{e}^{2}\alpha^{2}}{4p^{4}\sin^{4}(\frac{\theta}{2})},
\]
but deviations close to $\theta=0$ that are influenced by the choice
of $\epsilon.$

In eq. (\ref{eq:3.1}), the Coulomb phase shifts, $\sigma_{l}(p),$
have been replaced by the phase shifts of our model to second order.
Here
\begin{equation}
\delta=\frac{\beta T-2R}{\sigma_{x}}\label{eq:3.3}
\end{equation}
measures the time delay (positive) or advancement (negative) over
the time, $T,$ of the experiment, relative to free evolution. The
wavepacket centres are separated by $2R$ at the start of the experiment.
As discussed in \citet{Hoffmann2017a}, the choice $R/\sigma_{x}=\sqrt{\epsilon}$
gives an initial separation that grows as $\epsilon$ is decreased,
while wavepacket spreading is minimal over the course of the experiment.
The quantities
\begin{equation}
\Delta_{l}(p)=4\epsilon\eta\,\{\ln(4pR/e)-\psi(l+1)+\mathcal{O}(\eta)\},\label{eq:3.4}
\end{equation}
that influence this time shift involve the logarithmic phase, $-\eta\ln(2pr),$
from $\delta_{l}^{(1)}(p),$ which has been removed in the definition
\begin{equation}
\bar{\delta}_{l}^{(1)}(p)=\delta_{l}^{(1)}(p)+\eta\ln(2pr)=\eta\,\psi(l+1).\label{eq:3.5}
\end{equation}
The approximation in eq. (\ref{eq:3.4}) will be sufficient for our
purposes, as we will mostly be considering small $\eta.$

In what follows, we evaluate the phase shifts numerically and insert
them into the cross section formula, sum the series numerically and
take the modulus-squared.

\section{Numerical results}

\begin{figure}
\begin{centering}
\includegraphics[width=14cm]{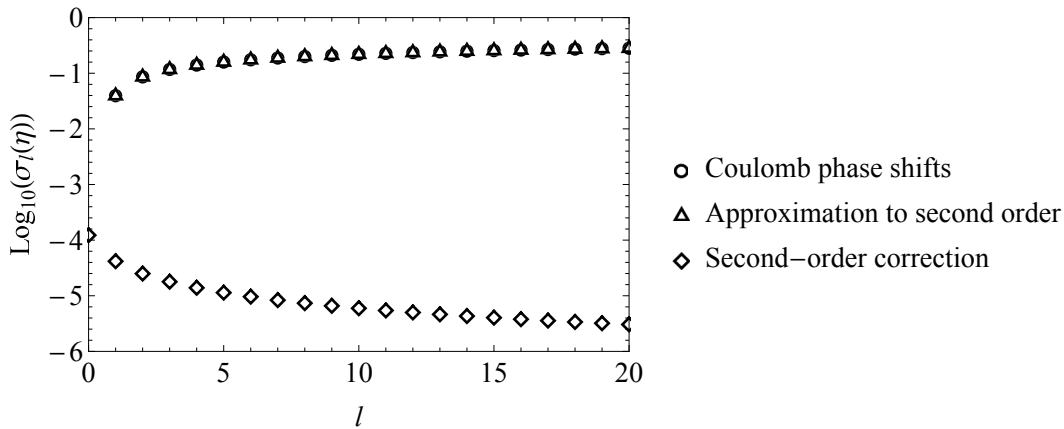}
\par\end{centering}
\caption{\label{fig:Comparison-of-phase}Comparison of exact and computed phase
shifts.}

\end{figure}

First we calculate the phase shifts and compare them to the Coulomb
phase shifts in the nonrelativistic regime ($p=0.02\,\mathrm{MeV}$,
$\beta=0.04$, $\eta=0.09$), where we expect agreement. We find the
results plotted in fig. \ref{fig:Comparison-of-phase}. The second-order
corrections are seen to be much smaller than the Coulomb phase shifts.
Note that the exact and approximate phase shifts are negative at $l=0$,
so are not included in this logarithmic plot.

That the second-order corrections were calculable and finite achieves
the first goal of this paper. That the corrections were small is as
it should be, since the first-order result gave a good approximation
to the Coulomb phase shifts.

As could be expected from this result, the plots of the model differential
cross section and the Rutherford formula were visually indistinguishable,
except close to $\theta=0,$ for this nonrelativistic choice of momentum.
We took $\delta_{l}^{(2)}(p),$ small but slowly varying, up to $l=50.$
The profile in $\delta$ at $\theta=90\lyxmathsym{\textdegree}$ peaked
very near $\delta=0.$

\begin{figure}
\begin{centering}
\includegraphics[width=14cm]{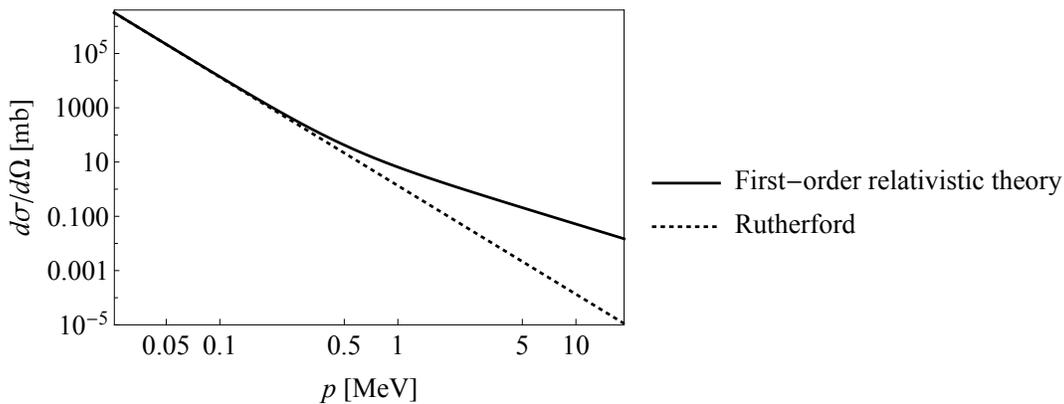}
\par\end{centering}
\caption{\label{fig:Model-differential-cross}Model differential cross section
at $\theta=90\lyxmathsym{\protect\textdegree}$ for a range of nonrelativistic
and relativistic momenta.}

\end{figure}

Into the relativistic regime, we plotted the model differential cross
section (just to first order) at $\theta=90\lyxmathsym{\textdegree}$
for a range of momenta from about $0.03\,\mathrm{MeV}$ to $19\,\mathrm{MeV}$
in fig. (\ref{fig:Model-differential-cross}), compared with the Rutherford
differential cross section.

We see agreement with the Rutherford differential cross section at
low momenta and a distinct deviation at higher momenta. One reason
that the relativistic result would be larger than the Rutherford result
is that the coupling strength in the former can be no lower than $\eta>\alpha/2,$
since $\beta<1,$ while the nonrelativistic form $\eta_{\mathrm{NR}}=\alpha m_{e}/p$
has a lower bound of zero.

\begin{figure}
\begin{centering}
\includegraphics[width=14cm]{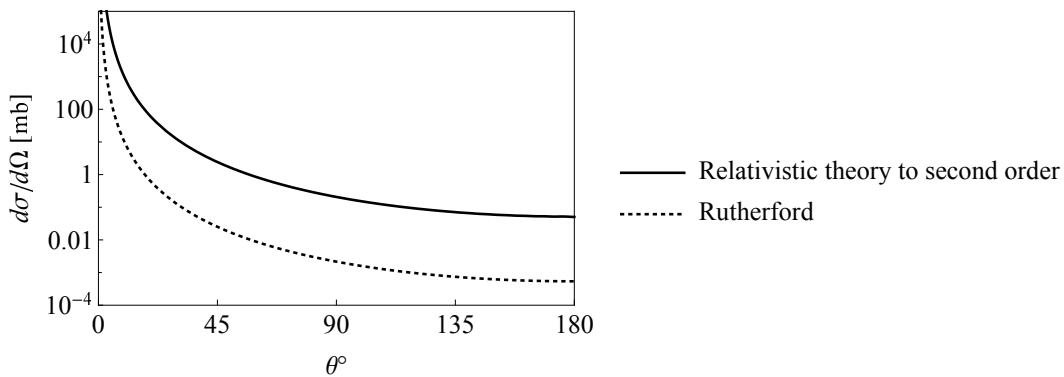}
\par\end{centering}
\caption{\label{fig:Angular-dependence-at}Angular dependence at $p=5\,\mathrm{MeV}.$}

\end{figure}

At $p=5\,\mathrm{MeV}$ ($\eta=0.004$) we plotted the differential
cross section as a function of angle, again comparing with the Rutherford
result. This is shown in fig. (\ref{fig:Angular-dependence-at}).
At $\theta=90\lyxmathsym{\textdegree},$ the $\delta$ profile peaked
very near $\delta=0.$ We note that the model cross section was seen
to take a large but finite value at $\theta=0,$ while the Rutherford
formula diverges there. This finiteness was discussed in \citet{Hoffmann2017a}.

At higher momenta, a more meaningful comparison is with the Møller
formula (\citet{Itzykson1980}). This is the differential cross section
derived from tree-level quantum electrodynamics for the scattering
of two electrons. To imitate a spinless system, it has been averaged
over initial spins and summed over final spins. The formula is
\[
\frac{d\sigma}{d\Omega}_{\mathrm{M\slashed{o}ller}}=\frac{1}{m_{e}^{2}}\frac{\alpha^{2}(1+\beta^{2})(1-\beta^{4})}{4\beta^{4}}\left[\frac{4}{\sin^{4}\theta}-\frac{3}{\sin^{2}\theta}\right]+\frac{1}{m_{e}^{2}}\frac{\alpha^{2}(1-\beta^{2})}{4}\left[1+\frac{4}{\sin^{2}\theta}\right].
\]
To make the comparison, we could not antisymmetrize these two fictitious
bosons. Instead, to have a definite particle exchange symmetry, we
symmetrized our model scattering amplitude, treating the two particles
as identical. This involved inserting a factor $1+(-)^{l}$ in the
sum over $l$ that defines the scattering amplitude.

\begin{figure}
\begin{centering}
\includegraphics[width=14cm]{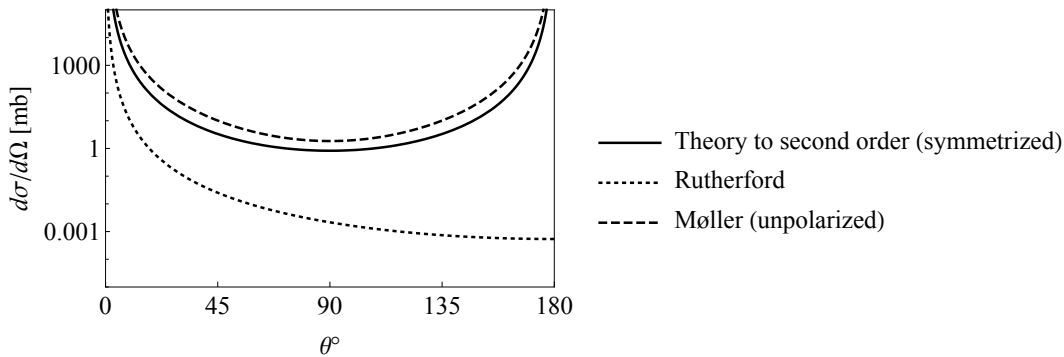}
\par\end{centering}
\caption{\label{fig:Angular-dependence-at-1}Angular dependence at $p=5\,\mathrm{MeV},$
compared with the Møller formula.}

\end{figure}

For $p=5\,\mathrm{MeV},$ we plotted the angular dependence of the
model cross section along with that of the Møller formula and the
Rutherford formula, in fig. (\ref{fig:Angular-dependence-at-1}).

\begin{figure}
\begin{centering}
\includegraphics[width=14cm]{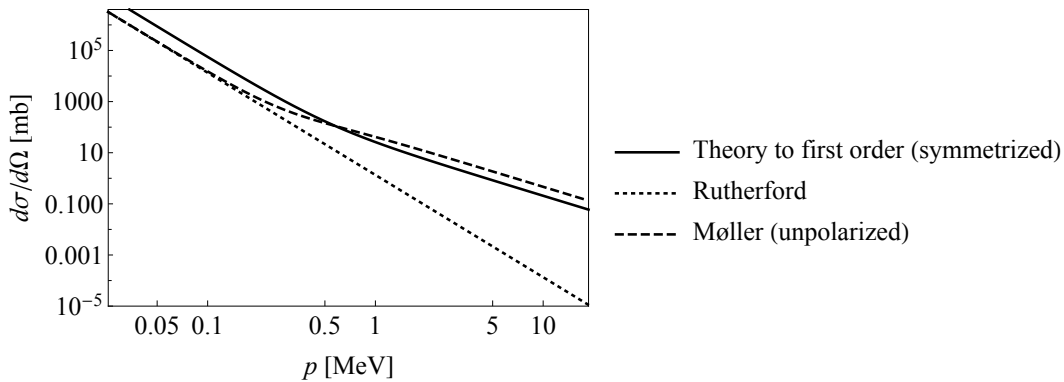}
\par\end{centering}
\caption{\label{fig:Momentum-dependence-at}Momentum dependence at $\theta=90\lyxmathsym{\protect\textdegree},$
compared with the Møller formula.}

\end{figure}

For the scattering angle $\theta=90\lyxmathsym{\textdegree},$ we
plotted the dependence on momentum of the model cross section and
the Møller formula, in fig. (\ref{fig:Momentum-dependence-at}).

It is remarkable that our model lacks many of the essential features
of the electron-electron scattering problem, other than the Coulomb
potential, yet shows quite good agreement with the Møller formula.
Note that the symmetrized model cross section no longer converges
to the Rutherford formula at nonrelativistic energies. Instead, it
must converge to a symmetric cross section that evidently has some
features in common with the Rutherford formula. We recall that the
unsymmetrized cross section was seen to converge to the unsymmetrized
Rutherford formula. It is then puzzling why the Møller formula converges
to a cross section that has no definite particle exchange symmetry.

\section{Conclusions}

There were two aims of this paper. One was to use a novel perturbation
theory on the Coulomb potential and show that finite results could
be obtained. In comparison, other forms of perturbation theory find
infinite results at second and all higher orders. This aim was achieved
using only one rule that was imposed rather than derived: the requirement
that principal part integration be used on the momentum integrals.

The other aim was to show that it is possible to construct a theory
that is relativistic and quantum-mechanical, but not quantum field-theoretic
and not the wave mechanics of a single particle in a potential. This
was achieved by working in momentum space, where the relativistic
dependence of single-particle energy on momentum can be easily introduced.
We found that the three-vector part of the four-potential must vanish
in the centre of mass frame. That leaves an arbitrary zero component,
at first sight not constrained by relativity. The further success
of this method depends on being able to choose a CM-frame potential
that reproduces experimental results.

The model of a Coulomb potential in the centre of mass frame was not
intended to provide a full description of the scattering of two electrons.
It was intended only as a proof of principle for the method considered.
There are other terms that contribute to the scattering of two electrons.
In a future work, we will include those terms, with the expectation
of finite results.

Also in a future work, we will analyze the forms of the terms in our
perturbation theory, to see if it is possible to prove finiteness
at higher orders.

\bibliographystyle{apsrev4-1}

\begin{thebibliography}{11}%
\makeatletter
\providecommand \@ifxundefined [1]{%
 \@ifx{#1\undefined}
}%
\providecommand \@ifnum [1]{%
 \ifnum #1\expandafter \@firstoftwo
 \else \expandafter \@secondoftwo
 \fi
}%
\providecommand \@ifx [1]{%
 \ifx #1\expandafter \@firstoftwo
 \else \expandafter \@secondoftwo
 \fi
}%
\providecommand \natexlab [1]{#1}%
\providecommand \enquote  [1]{``#1''}%
\providecommand \bibnamefont  [1]{#1}%
\providecommand \bibfnamefont [1]{#1}%
\providecommand \citenamefont [1]{#1}%
\providecommand \href@noop [0]{\@secondoftwo}%
\providecommand \href [0]{\begingroup \@sanitize@url \@href}%
\providecommand \@href[1]{\@@startlink{#1}\@@href}%
\providecommand \@@href[1]{\endgroup#1\@@endlink}%
\providecommand \@sanitize@url [0]{\catcode `\\12\catcode `\$12\catcode
  `\&12\catcode `\#12\catcode `\^12\catcode `\_12\catcode `\%12\relax}%
\providecommand \@@startlink[1]{}%
\providecommand \@@endlink[0]{}%
\providecommand \url  [0]{\begingroup\@sanitize@url \@url }%
\providecommand \@url [1]{\endgroup\@href {#1}{\urlprefix }}%
\providecommand \urlprefix  [0]{URL }%
\providecommand \Eprint [0]{\href }%
\providecommand \doibase [0]{http://dx.doi.org/}%
\providecommand \selectlanguage [0]{\@gobble}%
\providecommand \bibinfo  [0]{\@secondoftwo}%
\providecommand \bibfield  [0]{\@secondoftwo}%
\providecommand \translation [1]{[#1]}%
\providecommand \BibitemOpen [0]{}%
\providecommand \bibitemStop [0]{}%
\providecommand \bibitemNoStop [0]{.\EOS\space}%
\providecommand \EOS [0]{\spacefactor3000\relax}%
\providecommand \BibitemShut  [1]{\csname bibitem#1\endcsname}%
\let\auto@bib@innerbib\@empty
\bibitem [{\citenamefont {Dalitz}(1951)}]{Dalitz1951}%
  \BibitemOpen
  \bibfield  {author} {\bibinfo {author} {\bibfnamefont {R.~H.}\ \bibnamefont
  {Dalitz}},\ }\href@noop {} {\bibfield  {journal} {\bibinfo  {journal} {Proc.
  Roy. Soc. Lond. A: Math., Phys. and Eng. Sci.}\ }\textbf {\bibinfo {volume}
  {206}},\ \bibinfo {pages} {509} (\bibinfo {year} {1951})}\BibitemShut
  {NoStop}%
\bibitem [{\citenamefont {Collas}(2021)}]{Collas2021}%
  \BibitemOpen
  \bibfield  {author} {\bibinfo {author} {\bibfnamefont {P.}~\bibnamefont
  {Collas}},\ }\href@noop {} {\bibfield  {journal} {\bibinfo  {journal}
  {arXiv:2102.13105}\ } (\bibinfo {year} {2021})}\BibitemShut {NoStop}%
\bibitem [{\citenamefont {Hoffmann}(2021)}]{Hoffmann2021a}%
  \BibitemOpen
  \bibfield  {author} {\bibinfo {author} {\bibfnamefont {S.~E.}\ \bibnamefont
  {Hoffmann}},\ }\href {\doibase 10.1063/5.0023630} {\bibfield  {journal}
  {\bibinfo  {journal} {J. Math. Phys.}\ }\textbf {\bibinfo {volume} {62}},\
  \bibinfo {pages} {032105} (\bibinfo {year} {2021})}\BibitemShut {NoStop}%
\bibitem [{\citenamefont {Foldy}\ and\ \citenamefont
  {Wouthuysen}(1950)}]{Foldy1950}%
  \BibitemOpen
  \bibfield  {author} {\bibinfo {author} {\bibfnamefont {L.~L.}\ \bibnamefont
  {Foldy}}\ and\ \bibinfo {author} {\bibfnamefont {S.~A.}\ \bibnamefont
  {Wouthuysen}},\ }\href@noop {} {\bibfield  {journal} {\bibinfo  {journal}
  {Phys. Rev.}\ }\textbf {\bibinfo {volume} {78}},\ \bibinfo {pages} {29}
  (\bibinfo {year} {1950})}\BibitemShut {NoStop}%
\bibitem [{\citenamefont {Weinberg}(2012)}]{Weinberg2012}%
  \BibitemOpen
  \bibfield  {author} {\bibinfo {author} {\bibfnamefont {S.}~\bibnamefont
  {Weinberg}},\ }\href@noop {} {\emph {\bibinfo {title} {The Quantum Theory of
  Fields}}},\ Vol.~\bibinfo {volume} {1}\ (\bibinfo  {publisher} {Cambridge
  University Press, N.Y.},\ \bibinfo {year} {2012})\BibitemShut {NoStop}%
\bibitem [{\citenamefont {Messiah}(1961)}]{Messiah1961}%
  \BibitemOpen
  \bibfield  {author} {\bibinfo {author} {\bibfnamefont {A.}~\bibnamefont
  {Messiah}},\ }\href@noop {} {\emph {\bibinfo {title} {Quantum Mechanics}}},\
  Vol.\ \bibinfo {volume} {1 and 2}\ (\bibinfo  {publisher} {North-Holland,
  Amsterdam and John Wiley and Sons, N.Y.},\ \bibinfo {year}
  {1961})\BibitemShut {NoStop}%
\bibitem [{\citenamefont {Olver}\ \emph {et~al.}(2020)\citenamefont {Olver}
  \emph {et~al.}}]{DLMF2020}%
  \BibitemOpen
  \bibfield  {author} {\bibinfo {author} {\bibfnamefont {F.~W.~J.}\
  \bibnamefont {Olver}} \emph {et~al.},\ }\href {http://dlmf.nist.gov/}
  {\enquote {\bibinfo {title} {{NIST Digital Library of Mathematical
  Functions}},}\ }\bibinfo {howpublished} {http://dlmf.nist.gov/, Release
  1.0.25 of 2019-12-15} (\bibinfo {year} {2020})\BibitemShut {NoStop}%
\bibitem [{\citenamefont {Gradsteyn}\ and\ \citenamefont
  {Ryzhik}(1980)}]{Gradsteyn1980}%
  \BibitemOpen
  \bibfield  {author} {\bibinfo {author} {\bibfnamefont {I.~S.}\ \bibnamefont
  {Gradsteyn}}\ and\ \bibinfo {author} {\bibfnamefont {I.~M.}\ \bibnamefont
  {Ryzhik}},\ }\href@noop {} {\emph {\bibinfo {title} {Tables of Integrals,
  Series and Products}}},\ \bibinfo {edition} {corrected and enlarged}\ ed.\
  (\bibinfo  {publisher} {Academic Press, Inc., San Diego, CA},\ \bibinfo
  {year} {1980})\BibitemShut {NoStop}%
\bibitem [{\citenamefont {Hoffmann}(2017)}]{Hoffmann2017a}%
  \BibitemOpen
  \bibfield  {author} {\bibinfo {author} {\bibfnamefont {S.~E.}\ \bibnamefont
  {Hoffmann}},\ }\href@noop {} {\bibfield  {journal} {\bibinfo  {journal} {J.
  Phys. B: At. Mol. Opt. Phys.}\ }\textbf {\bibinfo {volume} {50}},\ \bibinfo
  {pages} {215302} (\bibinfo {year} {2017})}\BibitemShut {NoStop}%
\bibitem [{\citenamefont {Rutherford}(1911)}]{Rutherford1911}%
  \BibitemOpen
  \bibfield  {author} {\bibinfo {author} {\bibfnamefont {E.}~\bibnamefont
  {Rutherford}},\ }\href@noop {} {\bibfield  {journal} {\bibinfo  {journal}
  {Phil. Mag.}\ }\textbf {\bibinfo {volume} {Series 6, vol. 21}},\ \bibinfo
  {pages} {669} (\bibinfo {year} {1911})}\BibitemShut {NoStop}%
\bibitem [{\citenamefont {Itzykson}\ and\ \citenamefont
  {Zuber}(1980)}]{Itzykson1980}%
  \BibitemOpen
  \bibfield  {author} {\bibinfo {author} {\bibfnamefont {C.}~\bibnamefont
  {Itzykson}}\ and\ \bibinfo {author} {\bibfnamefont {J.-B.}\ \bibnamefont
  {Zuber}},\ }\href@noop {} {\emph {\bibinfo {title} {Quantum Field Theory}}},\
  \bibinfo {edition} {1st}\ ed.\ (\bibinfo  {publisher} {McGraw-Hill Inc.},\
  \bibinfo {year} {1980})\BibitemShut {NoStop}%
\end{thebibliography}

\end{document}